# Interaction of fcc-Palladium nano-crystals with Hydrogen during PECVD growth of Carbon nanotubes

Wilfried Wunderlich, Masaki Tanemura,

Nagoya Institute of Technology, Department of Environmental Technology, 466-8555 Nagoya
E-mail: wunder@system.nitech.ac.jp, tanemura@system.nitech.ac.jp

**Abstract**

Using plasma-enhanced chemical vapor deposition (PECVD) with Acetylene and Ammoniac on Pd-specimens Carbon nanotubes (CNT) could be produced successfully. Two different devices are compared and the conditions for best growth conditions are explained. The detailed analysis of electron diffraction pattern obtained by transmissions electron microscopy (TEM) of as-grown specimen showed an expansion of the Pd lattice, which can be explained by the formation of fcc-$PdH_x$ for x=0...1. For x=1..2 the first investigation of hexagonal -$PdH_2$ is reported, which lattice spacing is independent on the hydrogen content. The amount of fcc-PdH and hcp-$PdH_2$ increases when the specimens are treated subsequently with Hydrogen. A growth model is provided.

**Introduction**

Since the discovery of carbon nano-tubes (CNT) [1], their processing made a large progress and the possibility of controlled growth of nanotubes stimulated possible applications, for example they can be used as microsize X-ray emitters [2-3], flat displays or STM-tips, or for Hydrogen storage. The plasma-enhanced chemical vapor deposition (PECVD) method [4] is one of the most efficient production methods of nanotubes or their closed variants graphite nano-Fibers (GNF). The hydrocarbon gas is decomposed at the metal-surface, in the case considered in this paper Pd, leading to excess carbon forming the carbon walls and excess hydrogen gas.

Usually it is considered, that Hydrogen is stored interstitially in the octahedral vacancies of fcc-Pd nano-crystalline material with the possibility of an anisotropic lattice expansion in [111] and [100] directions [5]. hydrogen-charging with a concentration of $PdH_{0.706}$ leads to an expansion of the lattice constant to 0.4049nm instead of 0.3906nm in bulk-Pd [6]. During the CNT growth using the PECVD process catalytically decomposition of the inlet-gases occurs and atomic hydrogen is released, which is partly stored in the metallic nano-crystals lying inside the CNT. Since the generated hydrogen is known to be stored in the metallic lattice of the Pd-nano-crystals toping the CNT, this composite material is considered for efficient hydrogen storage.

Studies of the influence of hydrogen on the metallic nano-crystals inside the CNT are still lacking. The measurement of the lattice expansion of the palladium nano-crystallites inside the CNT is the main theme of this paper.

**Experimental**

Graphite nano-tubes were grown by the PECVD technique on a 500μm thin wire of Pd, as described previously [2-4]. Two devices were tested, one with a large reaction chamber (about 3 liter, fig1 a), another with a large vacuum tank connected with a 0.2mm orifice to the small reaction chamber (about 0.3 liter, fig 1b). The specimens produced in the smaller reaction chamber showed better results, due to reduced velocity of the gas molecules. Furthermore, the specimen lies on –400V (fig 1b) compared to earth (fig1a), which attracts the $H^+$-ions towards the specimen and also lead to more stable plasma conditions. By heating the wire inductively (2V 5A) it is heated to about T=500°C, which was found to be the best condition for growth, up to 5-30nm long fibers were found. The optimum-mixing ratio of acetylene ($C_2H_2$) and ammoniac ($NH_3$) gases was found as 1:2 (fig. 2), while at 1:3 the growth of triangular shaped carbon tubes and at 1:1.5 flat carbon layers were observed. The optimum partial pressure is 1.2 Torr, at lower partial pressures the amount is too low, at higher pressures the tubes became too fat. The ex-situ hydrogen-charging experiments on grown GNF were performed at 1 Torr $H_2$ at 400°C. For subsequent TEM-observations the wire was fixed on a 3mm-disc. The specimens were characterized by TEM JEOL 3010FX. The diffraction patterns were analyzed using image simulations performed with the image simulation program EMS, which is well known for TEM-analysis.

**Results and Discussions**

The low magnification TEM micrograph in fig. 3 a) shows the Pd nano-crystal embedded on top of a carbon nanotubes. From the analysis of the corresponding diffraction pattern it is deduced that the facets at the top of the nano-crystal consist of (220) planes, indicating that these high-density planes have the lowest energy to the surrounding carbon layers. The inner surfaces of the palladium crystals towards the empty fiber are also mostly facetted with a long tail. In previous analysis it was found that during decomposition of the acetylene gas hydrogen is released, which penetrates into the nano-tubes as well as into the metallic nano-crystals and it is concluded, that the melting point of Pd is reduced due to both, the hydrogen-interaction as well as the nano-size. In this case the Pd nano-crystals have a round shape indicating also the melting during CNT formation due to Hydrogen interaction.

In order to clarify the influence of Hydrogen the Pd-CNT-samples were exposed to $H_2$-gas at 400°C for 20 min. The diffraction pattern shown in fig. 3 b) correspond both to (022) diffraction spots in the zone axis [110] and both in the same magnification, the upper micrograph before, the lower after Hydrogen treatment.. The distance between the corresponding (022) spots and the incident beam is reduced after the hydrogen treatment, while the ring-like diffraction pattern from the carbon nanotubes in both figures have the same size. The conscientious analysis showed that the distance in reciprocal space of the (002) spot to the incident beam is about 3.1% smaller in the upper fig. 3 a) than in the lower, indicating the 3.1% extension of the lattice constants due to the hydrogen treatment. This leads to an extension of the lattice spacing to 0.404nm, in good agreement to previous measurements [6]. About 10% of the CNT, however, show diffraction pattern, which cannot be identified as fcc, but only as hcp crystals, as shown in fig. 3c. The analysis of the diffraction pattern in fig. 3c leads to a z=[22.1] zone axis with a lattice spacing of a=0.250nm and c=0.408nm. This c/a ratio of c/a=1.63 corresponds to other hcp metals like Co. Careful analysis of different zone axis, showed that the hexagonal phase only possess these lattice constants, which are independent to the hydrogen content, as confirmed by analysis of many diffraction pattern.

The increase of the lattice constants in fcc Pd due to interstitial hydrogen atoms is shown in fig. 4). The dots refer to the experimental bulk-observations [6], the straight line for fcc is the linear interpolation assuming the validity of the well-accepted Vegard rule for interstitial alloying of foreign elements. The hydrogen can be stored in the fcc lattice until all octahedral positions are filled, which corresponds to a maximal composition PdH or x=1 in $PdH_x$. Reaching this concentration the lattice constants have increased by 20pm, a value, which corresponds well to the hydrogen radius usual fitted for ionic materials. The fcc-(111) spacing increases from 0.225nm to 0.233nm for fcc $PdH_x$ from x=0 to 0.7 and would be 0.250nm for x=1. When the hydrogen content increases beyond x=1 to 2, the hexagonal lattice is observed, which c- axis is a continuation of the fcc-a-axis and the hcp-a axis is a continuation of the fcc-[111] spacing. Why the Vegard rule is not working is the case of the hexagonal $PdH_2$ requires some additional research.

In the crystal structure of hexagonal $PdH_2$ two types of interstitial tetraeder positions are possible, whether its tip points upward or downward (fig. 6). If both of these tetraeder type positions are fully occupied, the composition is $PdH_2$. This hcp-$PdH_x$ (with $x$=1…2) lattice has the space group $P6_3/mmm$, in which the two possible tetraeder sites are occupied with a probability of 50% in the case of x=1 or fully occupied in the case of $x$=2 ($PdH_2$). The atomic position for fcc-PdH and hcp-$PdH_2$ are listed in table 1). The symmetry would be reduced to $P6_3$ mc, if only the upper or lower half of them would be occupied. This symmetry-reduced lattice, however, does not fit to the experimentally observed diffraction pattern.

The hydrogen necessary to form PdH can be easily explained by the growth mechanism for nanotubes (fig. 6). On atomic scale the growth can be divided into five steps, first the cracking of the double C=H acetylen bonds, then the cracking of their carbon-hydrogen bonds, the formations of intermolecular hydrogen bonds and of carbon bonds in the nanotube shape, and finally the growth of the nanotube into the long shape. For all of these steps the catalytic function of Palladium is required. The growth only occurs at the tip of the nanotube, since splitted tubes are never observed. The growth of the nanotubes terminates, when the tip of the Pd-nanocrystal is covered with a graphite mono-layer. Hence, it is likely, that the growth occurs at the carbon-palladium interface. The decomposition of the acetylene gas releases the carbon required for the CNT growth and also the hydrogen, which is partially released in the vacuum, partly stored in the empty part of the CNT, and partly stored in the palladium crystal. Also the bias voltage of –500V enhances this hydrogen diffusion into the tubes.

**Summary**

This study reports about TEM investigation of the lattice expansion of fcc-palladium and the first investigation of hexagonal palladium hydride, which was found inside of CNT. Further research is required to find the optimal growth conditions to increase the amount of hcp-$PdH_2$. Since this is the first report on hcp-$PdH_2$ it has to be clarified, whether the H-penetration and hcp $PdH_2$ formation requires the nanotube/Pd interface as a necessary carrier acting like a funnel. The result of this study has a great impact for further storage of Hydrogen in Palladium. If the hexagonal palladium hydride structure could be stabilized, it will become possible to store a larger amount of hydrogen into the Pd lattice than previously considered.

**Acknowledgement**

We gratefully acknowledge the fruitful discussions and support by Prof. em. Fumio Okuyama.

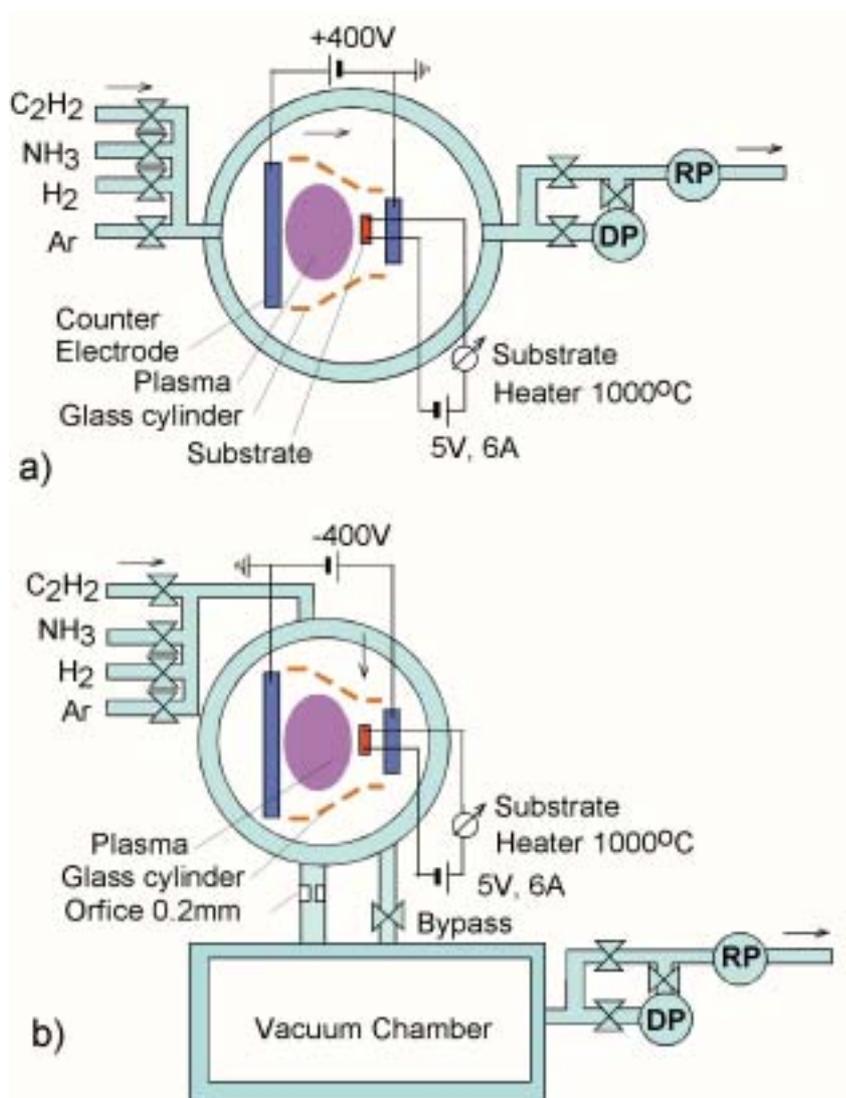

Fig. 1 Two different plasma-CVD equipments for producing Carbon nano tubes

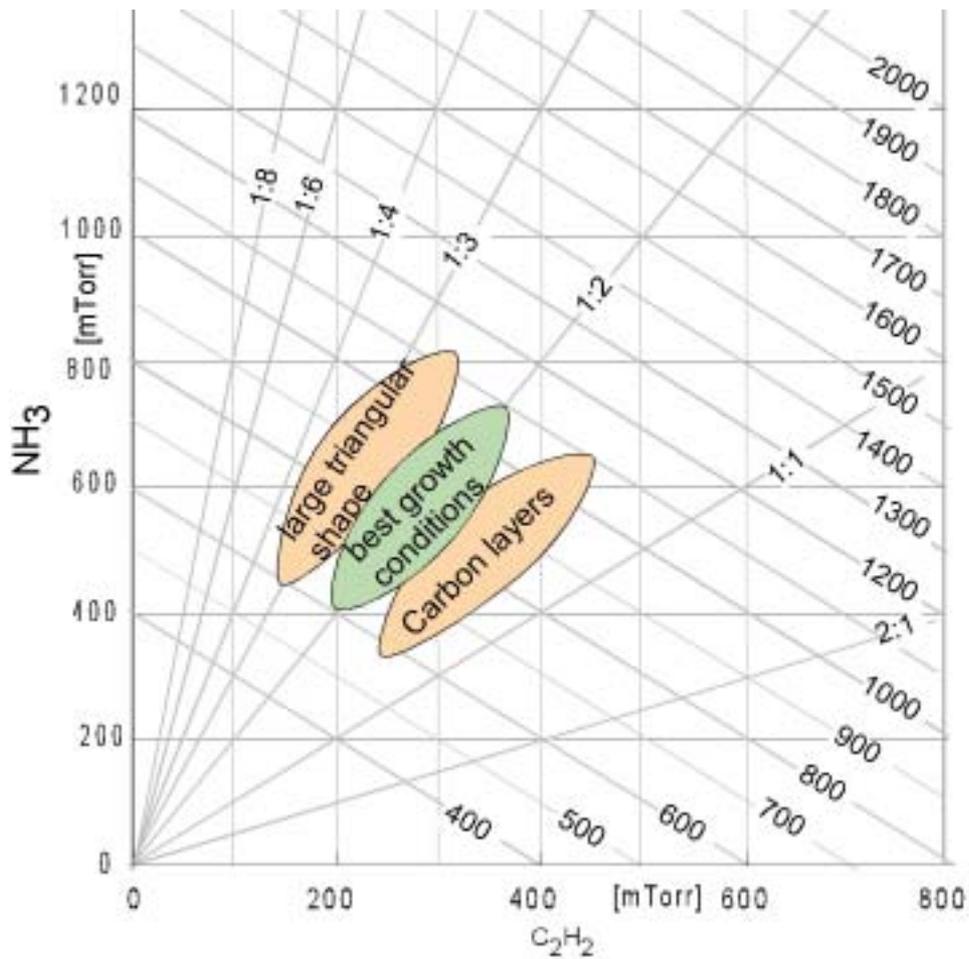

Fig. 2 The best CNT growth condition are found for the gas pressure ratio $P_{C2H2} : P_{NH3} = 1:2$

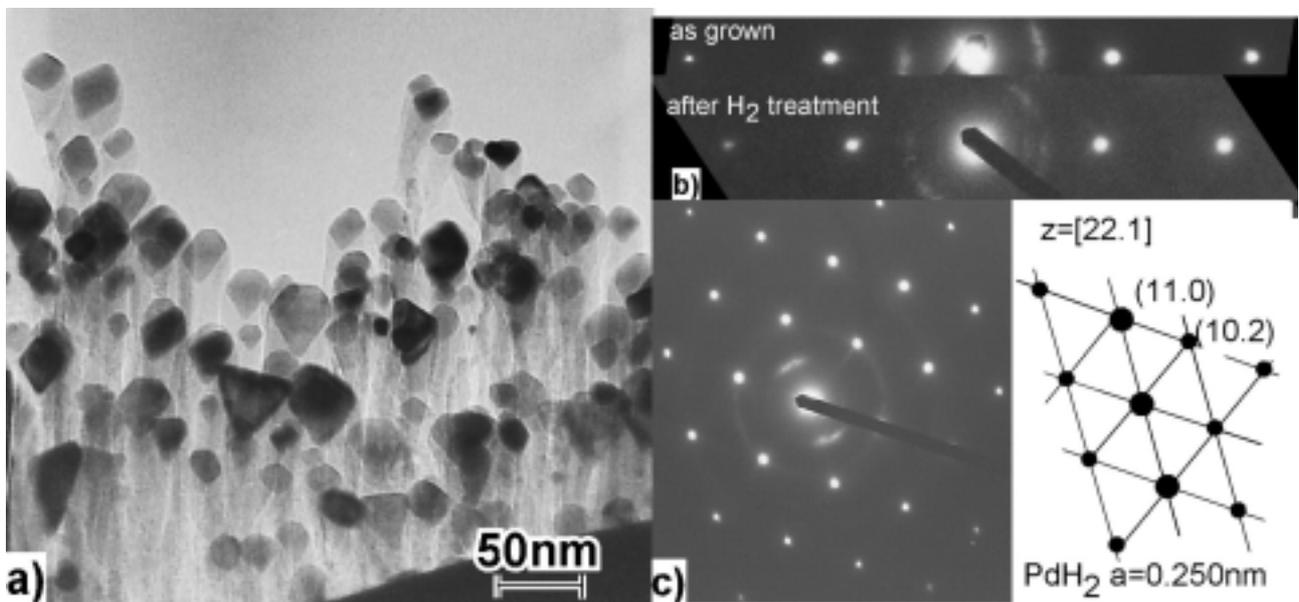

Fig. 3. TEM-observation of Palladium nano-crystals inside Carbon nanotubes, a) TEM-micrograph, b,c) Diffraction pattern b) with fcc structure, c) with hcp structure

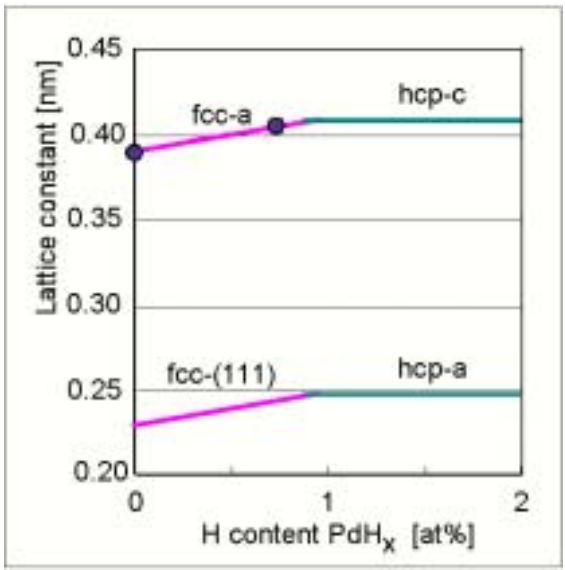

Fig. 4 Lattice constants for fcc PdH$_x$ (x=0…0.7) and hcp-PdH$_x$ (x=0.7…2) extrapolated from experimental data points [6] and deduced from TEM-diffraction pattern

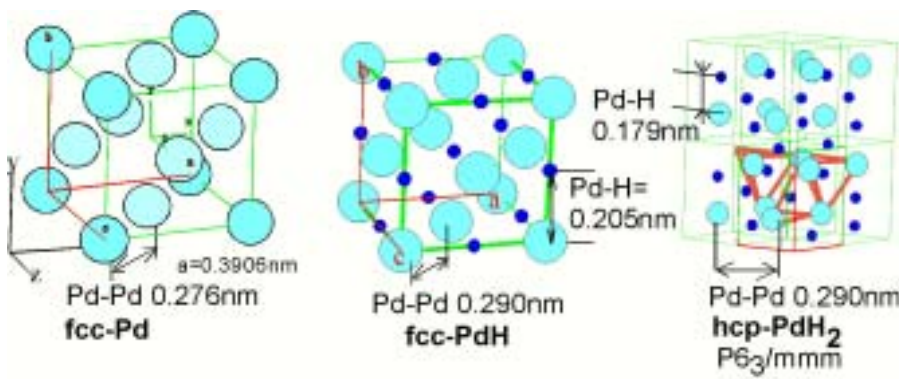

Fig.5 Crystal structure of a) fcc-Pd, b) fcc-PdH and c) hcp-PdH$_2$

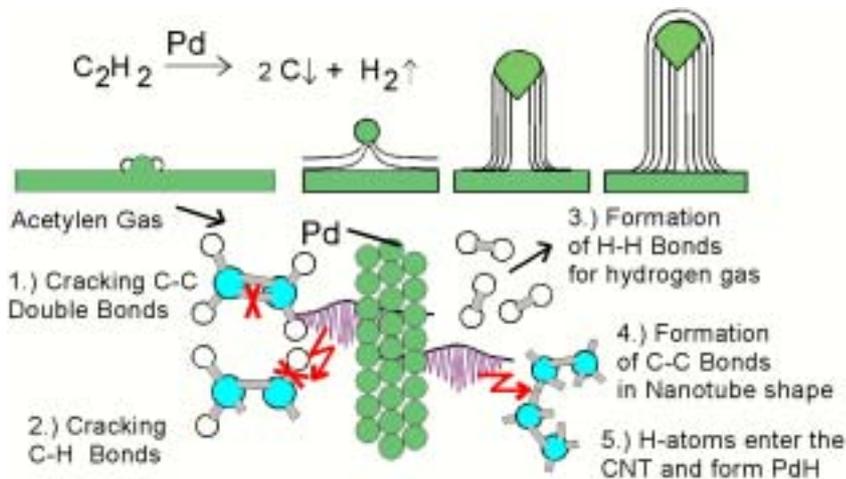

Fig. 6 Growth model for carbon nano tubes during plasma-CVD

6